\documentclass[print,pra,showpacs,preprintnumbers,twocolumn]{revtex4}
\usepackage{dcolumn}
\usepackage{amsmath}
\usepackage{bm}
\usepackage{graphicx}

\begin{document}

\title{Dissipative dynamics of quantum discord under quantum chaotic
environment}
\author{Y. Y. Xu$^{1,2}$, W. L. Yang$^{1}$}
\author{M. Feng$^{1}$}
\altaffiliation{Corresponding author: mangfeng@wipm.ac.cn}
\affiliation{$^{1}$State Key Laboratory of Magnetic Resonance and Atomic and Molecular
Physics, Wuhan Institute of Physics and Mathematics, Chinese Academy of
Sciences, Wuhan 430071, China}
\affiliation{$^{2}$Graduate School of the Chinese Academy of Sciences, Beijing 100039,
China}

\begin{abstract}
We investigate the dissipative dynamics of quantum discord in a decoherence
model with two initially entangled qubits in addition to a quantum kicked
top. The two qubits are uncoupled during the period of our study and one of
them interacts with the quantum kicked top. We find that the long time
behavior of quantum discord could be well described by the fidelity decay of
the quantum kicked top; for short time behavior, however, the phase of the
amplitude of the fidelity decay is necessary to provide more specific
information about the system. We have made comparison between the quantum
kicked top and multi-mode oscillator system in describing environment, and
also compared the dynamics of the entanglement with that of quantum discord.
\end{abstract}

\maketitle

It was a surprise that some quantum computing approaches could be carried
out efficiently in the absence of entanglement \cite{One}. This implies that
entanglement does not involve complete quantum correlation and thereby a new
terminology 'quantum discord' (QD) has drawn more and more attention \cite%
{Vedral,A1,Fan,wangb} recently and has been considered to replace
entanglement as a vital quantum resource for quantum information processing.

Since no quantum system is isolated from the environment, it is of
fundamental and practical importance to study the dissipative dynamics of
the QD. Modeling the environment by a multi-mode oscillator system \cite{FV}
had been widely adopted, but recent research pointed out that the
environment could also be simulated by the quantum chaos model \cite{QCM},
such as a quantum kicked top (QKT) \cite{epl}. It was shown that the quantum
chaos model could simplify the treatment about environment compared with the
multi-mode oscillator system model, and also efficiently simulate both
Markovian and non-Markovian environments \cite{Casati}.

Our focus in the present work is on the dynamics of the QD under a quantum
chaos model, in which the decoherence is from a QKT \cite{haake}, and the
decoherence could be reflected by fidelity decay (FD). We consider two
qubits with one of them coupled to a QKT. The two qubits are initially
entangled, but with no coupling during the period of our investigation. Our
purpose is to investigate the physics behind the dynamics of the QD under
this QKT-induced decoherence. To this end, we will check the dynamics of the
QD with respect to the chaotic and regular regimes of the QKT. We find that
FD is suitable for describing the long-term dynamics of the QD, which shows
difference between chaotic and regular QKTs. Moreover, from the phase of the
amplitude of FD (PAFD), we may discover more differences in the
environment-induced behavior between chaotic and regular QKTs. Furthermore,
we will try to explore the similarity, from the dynamics of the QD, between
the chaotic (regular) QKTs and Markovian (non-Markovian) multi-mode
oscillator environment. In addition, it has been shown that the relative
entropy entanglement (REE) is closely related to the QD \cite{Modi}. As a
result, a comparison of the QD with the REE would make us understand the
dissipative dynamics from another angle.

Let us first briefly review the main points of QD. The QD is defined to
quantify the difference of mutual information (MI) between the classical and
quantum forms. In classical information theory, the correlation between two
random variables $A$ and $B$ is described by MI using following two
expressions: $I(A,B)=H(A)+H(B)-H(A,B)$ and $J(A,B)=H(A)-H(\left.
A\right\vert B)$, where $H(A)$ and $H(B)$ are the Shannon entropies, $H(A,B)$
is the joint Shannon entropy, and $H(\left. A\right\vert B)$ is the
conditional entropy introduced for quantifying the ignorance about the value
of $A$ given a known $B$. The two expressions above are equivalent in the
classical case using Bayes' rule, and their quantum version is the
replacement of the Shannon entropy by von Neumann entropy, i.e., $I(\rho
_{A},\rho _{B})=S(\rho _{A})+S(\rho _{B})-S(\rho _{A,B})$, and $J(\rho
_{A},\rho _{B})=S(\rho _{A})-S(\left. \rho _{A}\right\vert \rho _{B})$,
where quantum conditional entropy $S(\left. \rho _{A}\right\vert\rho_{B})$
depends on the choice of measurement. If we perform a set of projectors \{$%
\Pi _{j}^{B}$\} on the subsystem $B$, then the post-measurement states of
the subsystem $A$ with outcomes $j$ is $\rho _{A|\Pi _{j}^{B}}=p_{j}Tr[\Pi
_{j}^{B}\rho _{AB}\Pi _{j}^{B}]$ with $p_{j}=Tr[I\otimes \Pi _{j}^{B}\rho
_{AB}]$, and the quantum conditional entropy is defined as $S(\left. \rho
_{A}\right\vert \rho _{B})=$ $\sum p_{j}S(\rho _{A|\Pi _{j}^{B}})$. So
different choices of \{$\Pi _{j}^{B}$\} result in different $S(\left. \rho
_{A}\right\vert \rho _{B})$, and the QD is defined as the minimum difference
of the quantum MI between $I(\rho _{A},\rho _{B})$ and $J(\rho _{A},\rho
_{B})$, i.e., $Q=\min [I(\rho _{A},\rho _{B})-J(\rho _{A}, \rho_{B})]$,
where $J(\rho _{A},\rho _{B})$ is also called classical correlation (CC)
\cite{Henderson}.

\begin{figure}[tbph]
\centering
\includegraphics[width=8cm]{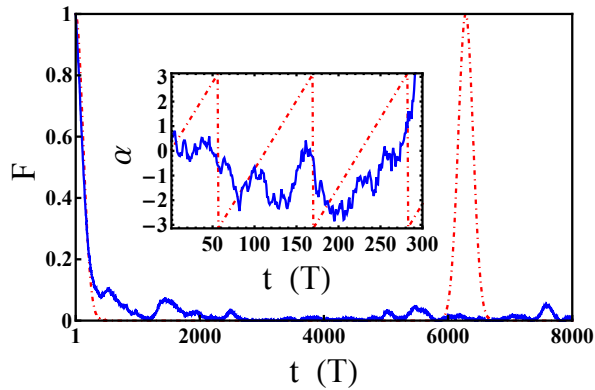}
\caption{(color online) Time evolution of the FD for chaotic (blue solid
line) and regular (red dot-dashed line) QKTs in units of T, where the
parameters are $J=100$, $\protect\nu =\protect\pi /2$ and $\protect\epsilon %
=0.001$. We set $\protect\eta =20$ and 0.1, respectively, for chaotic and
regular QKTs. The inset shows the time evolution of the PAFD $\protect\alpha
$ for chaotic (blue solid line) and regular (red dot-dashed line) QKTs in
short time scale, where, since the variation in the regular case is much
larger than that in the chaotic case, we enlarge $\protect\alpha $ by 40
times in the chaotic case, for a comparison with the regular case.}
\label{fig1}
\end{figure}

Consider the QKT with following Hamiltonian \cite{haake}
\begin{equation}
H_{E}=\frac{\nu }{T}J_{x}+\frac{\eta }{2J}J_{z}^{2}\sum\limits_{n=-\infty
}^{\infty }\delta (t-nT),
\end{equation}
where $J_{x}$ and $J_{z}$ are angular momentum components with $\hat{J}%
=J_{x} \hat{\imath}+J_{y}\hat{\jmath}+J_{z}\hat{k}$, $T$ is the time period,
and $\nu $ and $\eta $ represent the precessional angle and the kicking
strength, respectively. We assume that both the dynamics of the qubits and
dissipations are much slower than dephasing. So we may only focus on the
dephasing effect. We consider one of the qubits ($A$) is free and the other (%
$B$) interacts with QKT via the Hamiltonian
\begin{equation}
H=\varepsilon \sigma _{z}^{B}V_{E}+H_{E}
\end{equation}
where $\varepsilon $ is the strength of the interaction between the qubit
and the QKT, and $\sigma _{z}^{B}$ is the Pauli operator of the qubit B. $%
V_{E}$ is an operator of the QKT, which, for convenience, could be chosen as
$J_{x}$ in the following calculation. The corresponding Floquet operator can
be constructed as
\begin{equation}
U=e^{-i(\nu +\varepsilon \sigma _{z}^{B})J_{x}}e^{-i\frac{\eta }{2J}%
J_{z}^{2}}.
\end{equation}
In the classical limit $J\rightarrow +\infty $, the chaoticity degree is
determined by the kicking strength $\eta $ and the precessional angle $\nu $%
. For example, for $\nu =\pi /2$, the classical kicked top is dominated by a
regular motion for $\eta \leq 2.5$, whereas the motion is prevailed by chaos
for $\eta \geq 3$ \cite{F. Haake}. The regime with $2.5\leq \eta \leq 3$ is
for mixture of chaos and regularity. For our purpose in this work, we will
only consider the regular and chaotic regimes, instead of the mixture regime.

Consider following initial state of the system,
\begin{equation}
\rho _{ABE}=\frac{1}{4}(I+\sum_{j=x,y,z}c_{j}\sigma _{j}^{A}\otimes \sigma
_{j}^{B})\otimes \left\vert \theta ,\phi \right\rangle \left\langle \theta
,\phi \right\vert ,
\end{equation}
where $c_{j}$ are real numbers, and $\sigma _{j}^{A}$ ($\sigma _{j}^{B}$) is
the usual Pauli matrix acting on the subsystem $A$ $(B)$ \cite{Vedral}. The
initial QKT state $\left\vert \theta, \phi \right\rangle $ is a spin
coherent state defined as \cite{coherent state}
\begin{equation}
\left\vert \theta ,\phi \right\rangle =\exp [-i\theta (J_{x}\sin \phi
-J_{y}\cos \phi )]\left\vert J,J\right\rangle ,
\end{equation}
where $\left\vert J,J\right\rangle $ is the eigenstate of $\hat{J}^{2}$ and $%
J_{z}$ with eigenvalues $J(J+1)$ and $J$, respectively. The time evolution
of the density matrix after $n$ kicks can be described as
\begin{equation}
\rho _{ABE}^{\prime }=U^{n}\rho _{ABE}(U^{\dagger })^{n},
\end{equation}%
with the corresponding reduced density matrix of the two qubits, in the
basis $\{\left\vert 1\right\rangle =\left\vert 0\right\rangle_{A}\left\vert
0\right\rangle_{B},$ $\left\vert 2\right\rangle =\left\vert 0\right\rangle
_{A}\left\vert 1\right\rangle _{B},$ $\left\vert 3\right\rangle =\left\vert
1\right\rangle _{A}\left\vert 0\right\rangle _{B},$ $\left\vert
4\right\rangle =\left\vert 1\right\rangle _{A}\left\vert 1\right\rangle
_{B}\}$, given by
\begin{equation}
\rho _{AB}^{\prime }=\left(
\begin{array}{cccc}
\frac{1+c_{z}}{4} & 0 & 0 & \frac{(c_{x}-c_{y})f}{4} \\
0 & \frac{1-c_{z}}{4} & \frac{(c_{x}+c_{y})f}{4} & 0 \\
0 & \frac{(c_{x}+c_{y})f^{\ast }}{4} & \frac{1-c_{z}}{4} & 0 \\
\frac{(c_{x}-c_{y})f^{\ast }}{4} & 0 & 0 & \frac{1+c_{z}}{4}%
\end{array}
\right) ,
\end{equation}
where $f=\left\langle \Psi^{\prime }(t)|\Psi (t)\right\rangle =\left\langle
\theta ,\phi \right\vert \exp [i(\varepsilon V_{E}+H_{E})t]\exp
[-i(-\varepsilon V_{E}+H_{E})t]\left\vert \theta ,\phi \right\rangle $ is
the amplitude of FD of the QKT \cite{FD}, which is generally complex, i.e., $%
f=\sqrt{F}e^{i\alpha}$, where $F=\left\vert f\right\vert^{2}$ is the FD, and
the PAFD $\alpha$ reflects the relative phase between the QKT states $%
\left\vert \Psi (t)\right\rangle $ and $\left\vert \Psi ^{\prime
}(t)\right\rangle $. For a fully chaotic quantum system realized through
adjusting the parameter $\eta$, FD behaves with Gaussian or exponential
decay depending on the values of the perturbation strength \cite{FD}. Change
of the parameter $\eta $ moves the quantum system to the regular regime of
the QKT, and then yields only the Gaussian decay and revivals of the FD \cite%
{FD,FD revivals}. The corresponding revival time is $\tau =k/\varepsilon $
with $k$ the constant. Fig. 1 shows the FD for chaotic and regular QKTs. In
our calculation here and following, the time unit is the period $T$, and the
initial state $\left\vert\theta, \phi \right\rangle $ is a randomly chosen
spin coherent state. As shown in Fig. 1, the fluctuation in the chaotic case
could be understood as being from the finite numbers of the eigenstates of
the Floquet operator U involved in the initial QKT state $\left\vert \theta,
\phi \right\rangle$ \cite{Peres}, and the revivals in the regular case is
related to the underlying periodic classical motion \cite{FD revivals}. Both
the fluctuation and the revival could be regarded as the memory effects. The
inset in Fig. 1 shows the regular and chaotic behaviors reflected by PAFD:
Their variations are regular and irregular respectively, and the amplitude
of the variation in the regular case is much larger than that in the chaotic
case.

The reduced density matrix of the two qubits in Eq. (7) is an X state, whose
QD and CC could be calculated analytically, using the methods in \cite%
{Mazhar Ali}, as
\begin{equation}
Q(\rho _{A,B}^{\prime })=2+\sum\limits_{i=1}^{4}\lambda _{i}\log _{2}\lambda
_{i}-C(\rho _{A,B}^{\prime }),
\end{equation}
and
\begin{widetext}
\begin{equation}
C(\rho _{A,B}^{\prime })=1-\min
[-\sum\limits_{k=0}^{1}\frac{1+(-1)^{k}\theta _{1}}{2}\log
_{2}\frac{1+(-1)^{k}\theta _{1}}{2},-\sum\limits_{k=0}^{1}\frac{
1+(-1)^{k}\theta_{2}}{2}\log_{2}\frac{1+(-1)^{k}\theta_{2}}{2}],
\end{equation}
\end{widetext}where $\lambda _{1}=[1+c_{z}+\left\vert
(c_{x}-c_{y})\right\vert \sqrt{F}]/4$, $\lambda _{2}=[1+c_{z}-\left\vert
(c_{x}-c_{y})\right\vert \sqrt{F}]/4$, $\lambda _{3}=[1-c_{z}+\left\vert
(c_{x}+c_{y})\right\vert \sqrt{F}]/4$, $\lambda _{4}=[1-c_{z}-\left\vert
(c_{x}+c_{y})\right\vert \sqrt{F}]/4$, $\theta _{1}=\left\vert
c_{z}\right\vert $, and $\theta _{2}=\sqrt{[2(c_{x}^{2}+c_{y}^{2})+2\left%
\vert c_{x}^{2}-c_{y}^{2}\right\vert (\left\vert \cos 2\alpha \right\vert
+\left\vert \sin 2\alpha \right\vert )]F}/2$. For a real or imaginary
amplitude of the FD, $\theta _{2}$\ could be reduced to $\theta _{2}=\sqrt{F}%
\max [\left\vert c_{x}\right\vert ,\left\vert c_{y}\right\vert ]$, which is
similar to the result of the phase damping channel in \cite{Vedral},
describing the dephasing induced by the Markovian multi-mode oscillator
environment.

To be more clarified, we compare the QD with the REE for entanglement \cite%
{Knight}. Direct calculation yields the REE in our case,
\begin{equation}
E=1+\beta ^{\prime }\log _{2}\beta ^{\prime }+(1-\beta ^{\prime })\log
_{2}(1-\beta ^{\prime }),
\end{equation}%
where $\beta ^{\prime }=\max [0.5, \lambda _{\max }]$, with $\lambda _{\max
}=\max [\lambda_{1}, \lambda_{2}, \lambda_{3}, \lambda_{4}].$ We plot the
evolution of QD and REE, along with CC and FD, in Figs. 2 and 3 for a
comparison. In the chaotic regime, both the QD and REE are only determined
by the FD, but in the regular regime, the REE still only depends on the FD,
whereas QD behaves complicated: Its long time behavior is fully determined
by FD; the short time behavior is more relevant to the PAFD $\alpha$. $%
\alpha $.

\begin{figure}[tbph]
\centering
\includegraphics[width=8 cm]{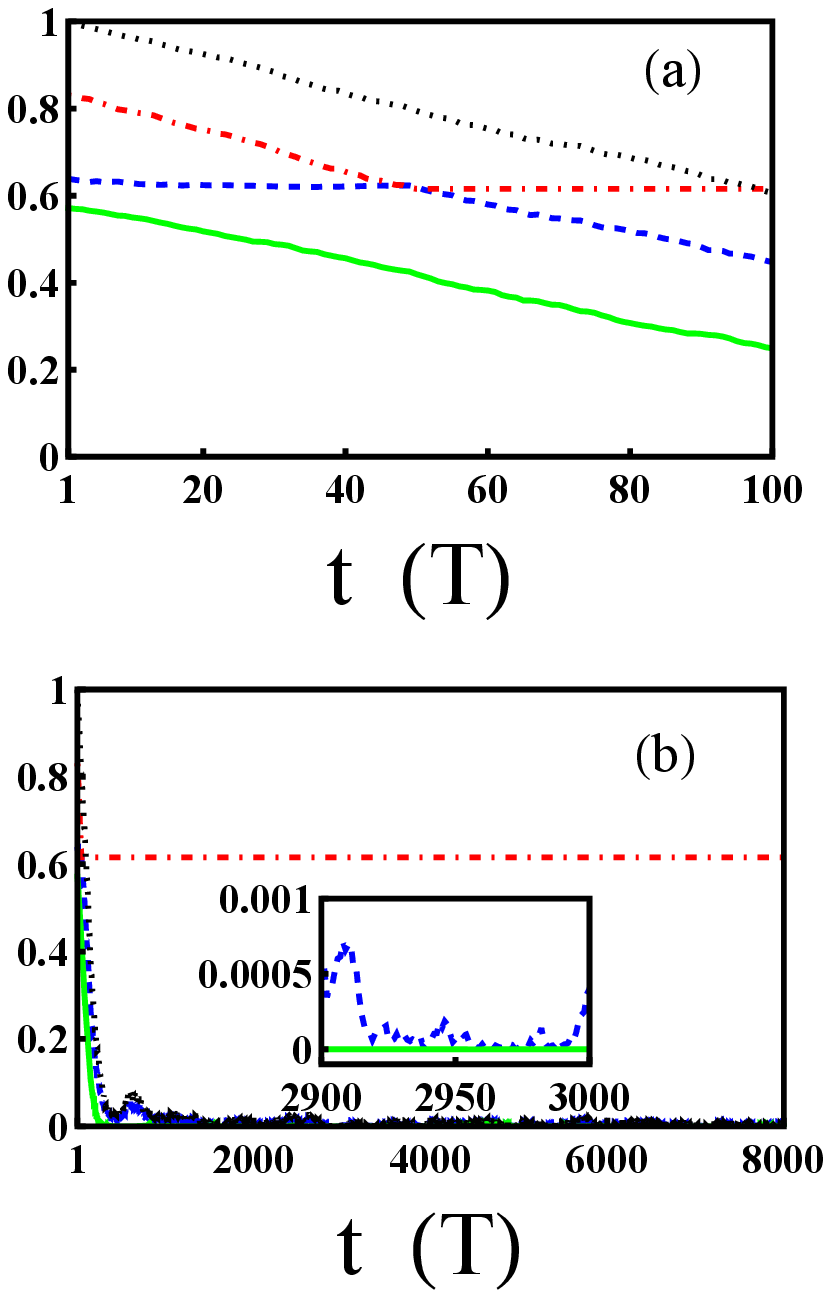}
\caption{(color online)(a) Short time evolution of the QD (blue dashed
line), CC (red dash dotted line), REE (green solid line), and FD (black
dotted line) in units of T with respect to the chaotic QKT, where the
parameters of the FD are the same as in Fig. 1 and the initial state is with
the parameters $c_{x}=$0.95, $c_{y}=$-0.85, and $c_{z}=$0.85. (b) Long time
evolution of the QD, CC, REE, and FD under the same parameters in (a). The
inset is a magnification for the time period $2900<t/T<3000$.}
\label{fig2}
\end{figure}

To further check the Markovian and non-Markovian dephasing effects of the
QKT, we consider below the correspondence associated with the chaotic and
regular QKTs, respectively. For the chaotic QKT, the PAFD $\alpha $ is
nearly zero during the short time evolution of the FD. So the effect from
the chaotic QKT is similar to the phase damping channel, a typical Markovian
environment. Like the model of the phase damping channel, there are three
kinds of CC dynamics in our case here, depending on the parameters $c_{j}$,
i.e., the constant CC, CC with a sudden change, and CC with monotonic decay
\cite{Vedral}. Our interest is in the case with a sudden change, where the
QD can be maintained for some time \cite{Mazzola}. To show similarities and
differences between our decoherence model and the phase damping channel, the
dynamics of QD, CC and others, whose short time behaviors are very similar
to those under the phase damping channel \cite{Mazzola}, are plotted in Fig.
2. But for long time behavior, the situation is different: For example,
there are fluctuations in the evolution of the QD for a finite angular
momentum $J$, which does not happen in the phase damping channel. In
addition, we could find from the Fig. 2 (a) the consistency in the short
time evolution between the CC and the FD, and also between the REE and the
FD. Furthermore, in comparison to the sudden death in the time evolution of
REE, there is only instantaneous disappearance of the QD at some time points
as shown in Fig. 2 (b), where the small fluctuations of the QD are
originated from the memory effects of the QKT. Those fluctuations disappear
only in the classical limit, i.e., $J\rightarrow +\infty$ in our case, which
corresponds strictly to the Markovian regime.

\begin{figure}[tbph]
\centering
\includegraphics[width=7.6 cm]{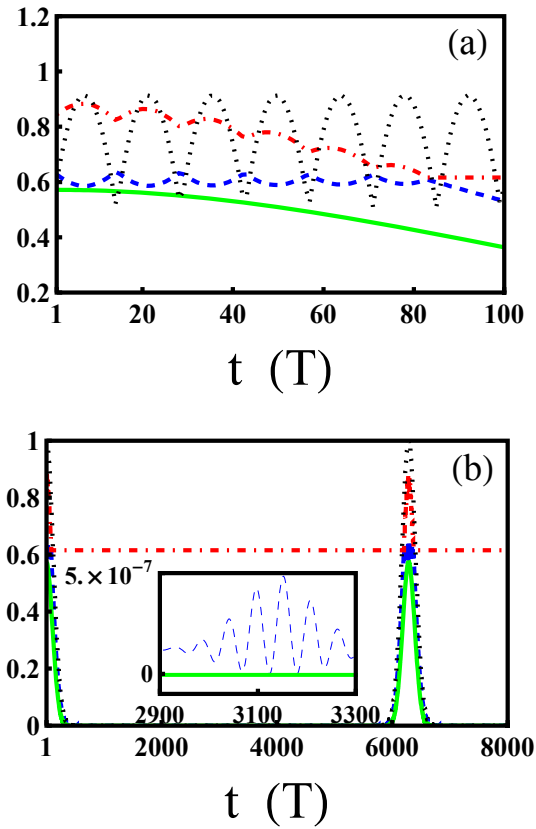}
\caption{(color online) (a) Short time evolution of the QD (blue dashed
line), CC (red dash dotted line), REE (green solid line), and the compressed
quantity $\left\vert \cos 2\protect\alpha \right\vert +\left\vert \sin 2%
\protect\alpha \right\vert $ (black dotted line) relevant to the regular QKT
in units of T, where other parameters are the same as in Fig. 2 except $%
\protect\eta =$0.1. (b) Long time evolution of the QD (red solid line), CC
(blue dashed line), REE (green solid line), and the quantity FD (black
dotted line) under the same parameters in (a). The inset is a magnification
for the time period $2900<t/T<3300$.}
\label{fig3}
\end{figure}

The memory effects of the QKT could be more evident in the treatment of the
regular QKT, which resembles the dynamics in the non-Markovian environment
\cite{wangb}. From short time behaviors, we may find that before the CC
becomes a constant, there exist oscillations in both the CC and the QD, as
shown in Fig. 3 (a). Although these oscillations are very similar to those
in a non-Markovian environment \cite{Fan}, their origins are different. The
oscillation in the non-Markovian environment is related to the dephasing
\cite{non markov}, whereas in our case it comes from the PAFD. In contrast,
there is no oscillation in the short time behavior of the REE (see Fig.
3(a)), which could be considered as an important difference between
entanglement and QD. Another difference between entanglement and QD could be
found in the long time evolution (See Fig. 3(b)), where REE experiences
sudden death and birth, but not for QD.

Therefore, only the chaotic QKT in the classical limit describes the
Markovian environment. The rest of the QKT, such as the finite J of the
chaotic QKT as well as the regular QKT, corresponds to the non-Markovian
environment. In this sense, the results in \cite{Mazzola} are only relevant
to the chaotic QKT in classical limit. It would be of great interests to
revisit the problems in \cite{Mazzola} by regular QKT or finite J of chaotic
QKT. Moreover, entanglement is usually described by concurrence. If
concurrence is employed in our model, however, the main results obtained
above would be remained, because both REE and concurrence involve no PAFD.
So entanglement could be distinguished from QD by PAFD regarding the regular
regime of the environment.

Our model could be straightforwardly extended to some more complicated
cases, such as two qubits interacting with a QKT commonly or with two QKTs
independently, particularly from the initial state Eq. (4). As a result, our
model would be useful to study either two locally located quibts
experiencing the same environment or two distant qubits suffering different
decoherence. Moreover, our model involves only dephasing induced by the QKT.
It might be more interesting to take into account a model involving more
complicated dissipation in the future.

In summary, we have studied the dissipative dynamics of the QD for two
qubits initially entangled, but only one of them coupled by a QKT. We have
checked the dynamics of the QD in both the chaotic and regular regimes of
the QKT, in comparison to entanglement. A comparison of the dynamics of the
QD under QKT with the multi-mode oscillator model has also been made, which
would help to go beyond multi-mode oscillator approximation and to
understand dissipative dynamics from another viewpoint. Since QD could be
used to understand the efficiency of quantum gate operations in the absence
of entanglement and quantum systems might suffer from quantum chaotic
environment, our study should be helpful for not only further understanding
quantum chaos, but also suppressing decoherence in quantum information
processing.

This work is supported by National Natural Science Foundation of China under
Grant No. 10774042.

\end{document}